\documentclass[10pt,twocolumn,
superscriptaddress,
 amsmath,amssymb,
 aps,
pra,
]{revtex4-2}
\usepackage{graphicx}
\usepackage{dcolumn}
\usepackage{bm}
\usepackage{xcolor}
\usepackage{soul}
\usepackage[colorlinks,linkcolor=blue,anchorcolor=blue,citecolor=blue,]{hyperref}
\newcolumntype{.}{D{.}{.}{-1}}
\newcolumntype{d}[1]{D{.}{.}{#1}}
\newcommand*{\wn}{cm$^{-1}$}

\renewcommand{\eqref}[1]{Eq.~(\ref{#1})}

\begin{document}

\title{Letokhov-Chebotayev intracavity trapping spectroscopy of H$_2$} 

\author{Wim Ubachs}
 \affiliation{Department of Physics and Astronomy, LaserLaB, Vrije Universiteit \\
 De Boelelaan 1100, 1081 HZ Amsterdam, The Netherlands}
  \author{Frank M.J. Cozijn}
  \affiliation{Department of Physics and Astronomy, LaserLaB, Vrije Universiteit \\
 De Boelelaan 1100, 1081 HZ Amsterdam, The Netherlands}
 \author{Meissa L. Diouf}%
 \affiliation{Department of Physics and Astronomy, LaserLaB, Vrije Universiteit \\
 De Boelelaan 1100, 1081 HZ Amsterdam, The Netherlands}
  \author{Clement Lauzin}
 \affiliation{Institute of Condensed Matter and Nanosciences, Université catholique de Louvain, \\ Louvain-la-Neuve, Belgium}
 \author{Hubert J\'{o}\'{z}wiak}
 \affiliation{Institute of Physics, Faculty of Physics, Astronomy and Informatics, Nicolaus Copernicus University in Toru\'{n}, \\ 
 Grudziadzka 5, 87-100 Toru\'{n}, Poland}
  \author{Piotr Wcis\l{}o}
 \affiliation{Institute of Physics, Faculty of Physics, Astronomy and Informatics, Nicolaus Copernicus University in Toru\'{n}, \\ 
 Grudziadzka 5, 87-100 Toru\'{n}, Poland}

\date{\today}

\begin{abstract}
\noindent
In the early days of laser spectroscopy Letokhov and Chebotayev proposed a scheme for measuring narrow spectral lines where the resolution is not restricted to Doppler effects because the molecules are entrained in a standing-wave light field. 
Now,  such one-dimensional trapping 
in the intensity maxima of an intracavity field, slightly detuned from resonance, is experimentally demonstrated in the measurement of the very weak S(0) (2-0) quadrupole overtone transition in H$_2$ at 1189 nm. 
The trapping manifests as an extremely narrow absorption feature at the predicted zero-recoil position, a 70 kHz shift from the blue-recoil component observed in Lamb-dip spectroscopy. A quantitative analysis of the saturation and trapping conditions supports the findings.

\end{abstract}

\maketitle

Developments in precision frequency measurements of atoms and molecules have relied on methods to overcome the Doppler effect resulting from the motion of the absorbing particles. 
Molecular beams are a method of choice for reducing the velocity components along the light vector leaving only a small residual effect~\cite{Ramsey1950}. 
The limitations of the Doppler effect become canceled, at least in first order, under conditions of Lamb-dip  saturation~\cite{Szoke1963} or in coherent two-photon absorption with counter-propagating laser beams~\cite{Levenson1974}. 
Still, molecular velocities affect the spectroscopy via transit-time broadening, connected to the Heisenberg uncertainty principle~\cite{Demtroeder}.
Trapping of particles, either for ions in oscillating electromagnetic fields~\cite{Paul1990}, via optical dipole trapping~\cite{Ashkin1970}, or via dissipative laser cooling techniques~\cite{Neuhauser1978}, mitigates this velocity-dependent broadening mechanism.
In strong fields the translational energy becomes quantized and trapping can occur in a ladder of states, between which optical sideband cooling or anti-Stokes Raman pumping can occur~\cite{Wineland1978}.
The laser-based cooling and trapping methods, initially pursued for atoms, are currently applied to molecules, whereby their more complex level structure poses various challenges to be overcome by novel approaches~\cite{Norrgard2016,Ding2020,Vilas2022}.

Letokhov and Chebotayev described a potential scheme of intracavity saturated absorption~\cite{LetokhovChebotayev}, in essence already put forward by one of these authors in 1968~\cite{Letokhov1968}, whereby the Doppler effect is avoided by trapping of molecules in a standing-wave light field.
That should be possible for particles with very low absolute speed, or for particles that move at very narrow angles to the wave front~\cite{LetokhovChebotayev}.

\begin{figure}[hb]
\begin{center}
\includegraphics[width=1\linewidth]{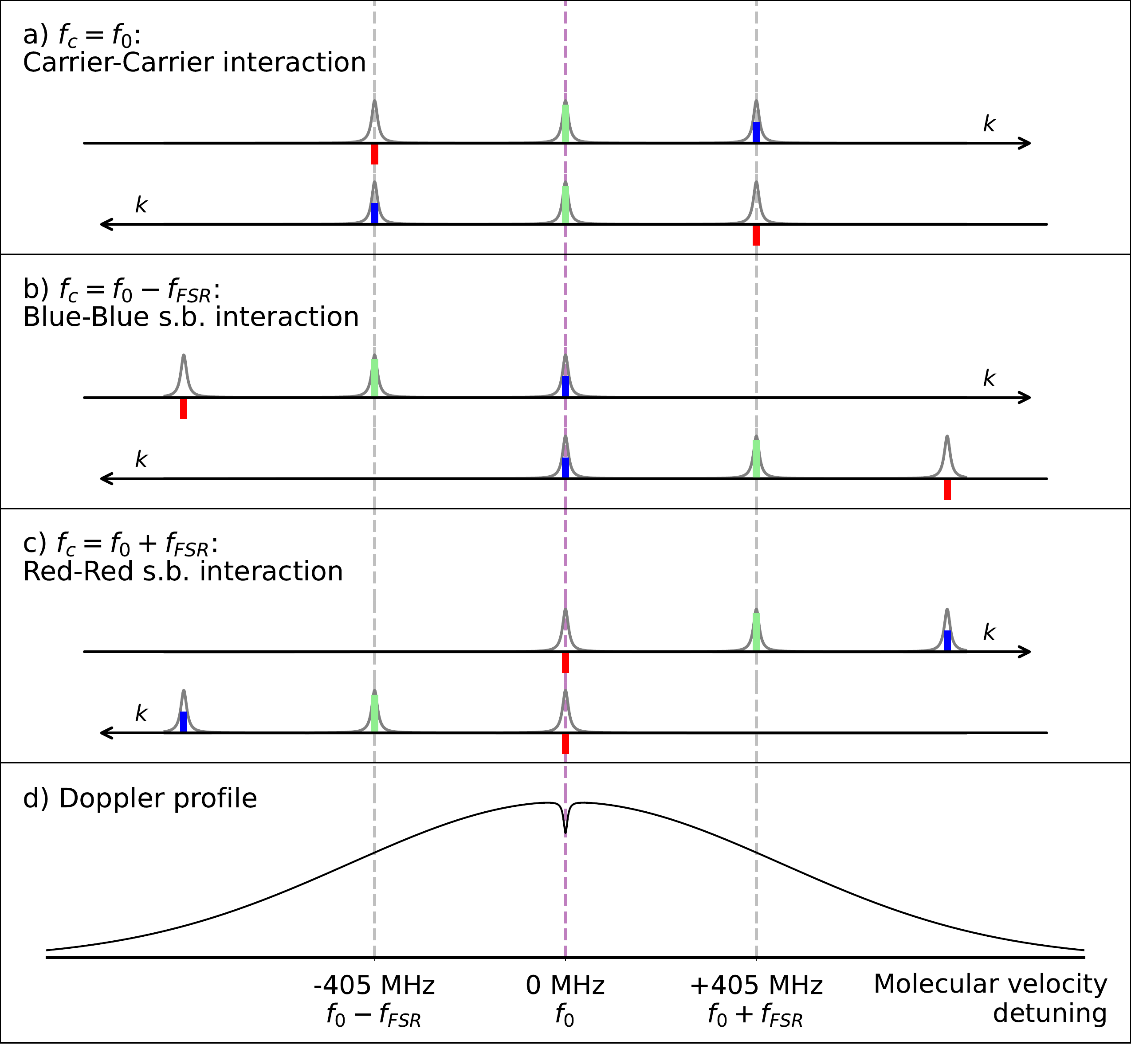}
\caption{\label{Sideband-scheme}
The three optical configurations used for probing the $v_y=0$ velocity class 
of the S(0) (2-0) quadrupole line of H$_2$ at resonance by the three counter propagating fields.
(a) Carrier-Carrier scheme; (b) Blue-Blue sideband scheme; (c) Red-Red sideband scheme; (d) the Doppler profile of H$_2$ at 72 K with a representation of a Lamb dip.}
\end{center}
\end{figure}

The present study reports on an experimental demonstration of the proposed one-dimensional (1D) trapping scheme, taking advantage of an intra-cavity configuration combined with the noise-immune cavity-enhanced optical heterodyne molecular spectroscopy (NICE-OHMS) technique~\cite{Ye1996} delivering superior detection sensitivity.
It builds further on the precision metrology experiment on the extremely weak S(0) quadrupole transition in the (2-0) overtone band of H$_2$~\cite{Cozijn2023}, corresponding to the transition from $J=0$ in $v=0$ to $J=2$ in $v=2$, and use this as a benchmark.
The typical NICE-OHMS opto-electronic configuration involves a carrier frequency $f_c$ and two opposite-phased sidebands at $f_c \pm f_{\rm FSR}$ (with FSR the free spectral range of the cavity) which are copropagating inside an optical resonator~\cite{Axner2008}.
The modulation and demodulation of side-bands in NICE-OHMS usually serve the purpose of shifting the ($1/f$) noise floor to higher frequencies, therewith enhancing the sensitivity for absorption detection, but in the present study the three frequency fields serve various purposes. 

In Fig.~\ref{Sideband-scheme} different optical excitation schemes are presented using the three radiation fields at frequencies $f_c$ and $f_c \pm f_{\rm FSR}$. 
Three schemes are employed to record a saturated Lamb-dip of the S(0) (2-0) transition in H$_2$ under selection of molecules at a velocity component $v_y=0$ along the laser propagation direction in the cavity (defined as the $y$-axis).

The carrier-carrier scheme was employed previously~\cite{Cozijn2023} and delivered a transition frequency of $f = 252\,016\,361\,234$ (8) kHz in extrapolation to zero pressure and power level.
The measured linewidth, observed as a symmetric line represented by a Lorentzian profile of 250 kHz (FWHM), is mainly determined from transit-time broadening at cryogenic temperature (72 K), with a small contribution by pressure broadening.
In the scheme denoted as blue-blue-sideband the H$_2$ absorption line is probed not by the carrier wave in the cavity, but by the blue-sideband. 
For this purpose the laser (carrier) frequency is shifted by one FSR ($405$ MHz) to lower frequency.  
For the red-red-sideband scheme the opposite is the case. 
The carrier frequency is measured via a beatnote to the output of a frequency-comb laser, that is locked to a Cs atomic clock, while the cavity is Pound-Drever-Hall (PDH)-locked to the comb~\cite{Drever1983}, and the modulation frequency is locked to the FSR via a DeVoe-Brewer scheme~\cite{Devoe1984} as was documented previously~\cite{Cozijn2023,Tobias2020}.

\begin{figure}[htb]
\begin{center}
\includegraphics[width=1.0\linewidth]{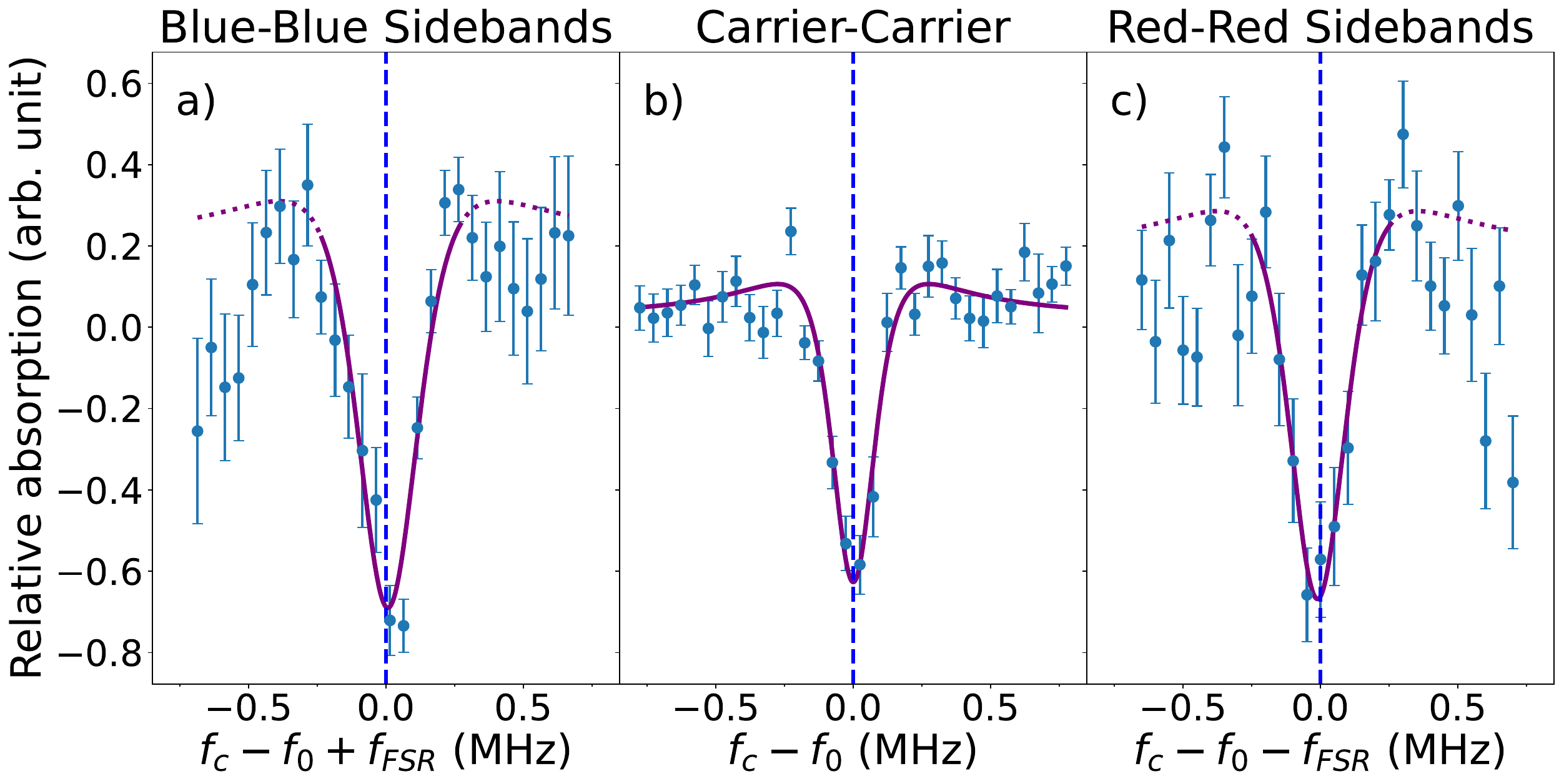}
\caption{\label{Lambdips}
Saturated absorption (Lamb dip) spectra recorded under application of the three optical configurations as displayed in Fig.~\ref{Sideband-scheme} at $p=$ 0.5 Pa and $T=72$ K. All three measurements were conducted with 150 W of saturation power, but for the sideband measurements a 2 kW strong detuned carrier wave was also present. The experimental data is fitted with a standard 1st derivative of a dispersive Lorentzian. For the sideband measurements the fit is only constrained in the center (solid line) as the lineshape starts to deviate at the wings, possibly due to effects of the strong carrier wave. Carrier (b), blue-blue-sideband (a), and red-red sideband (c) schemes indicated at the tops. The dashed (blue) vertical line in each of the panels refers to the same $f_0 = 252\,016\,361\,229$ kHz frequency, which is offset by the known modulation frequency $f_{\rm FSR}$, and  corresponds to the blue-recoil position (see text). 
}
\end{center}
\end{figure}

The three optical schemes displayed in Fig.~\ref{Sideband-scheme} are applied to record the S(0) (2-0) Lamb dip. Each of the spectra was recorded with  150 W of circulating power on resonance (at beam waist radius $\omega_0=0.54$ mm), either in the carrier, or in the sidebands. 
Due to the extreme weakness of this overtone quadrupole transition ($A=1.3 \times 10^{-7}$ s$^{-1}$) it took about 4 hours of averaging to record the carrier-carrier spectrum (and 8 hours for the sideband excitation spectra) in saturation. Resulting spectra are shown in Fig.~\ref{Lambdips}.
Important to note for the following discussions, in the cases of sideband excitation (blue-blue and red-red) there is an off-resonant carrier beam present at the 2 kW power level.

The Lamb-dip in the carrier-carrier spectrum, in Fig.~\ref{Lambdips}(b), was here measured at a pressure of 0.5~Pa resulting in an observed resonance frequency at $f~=~252\,016\,361\,229$ (6) kHz, in good agreement with the extrapolated value reported in Ref.~\cite{Cozijn2023}, for a pressure shift of -15 kHz/Pa and uncorrected for the recoil shift.
This position is indicated in all spectra with the blue-(dashed) vertical line, in which the position is shifted with the modulation frequency $f_{\rm FSR}$ for the sideband spectra.
In the red-red sideband spectrum, in Fig.~\ref{Lambdips}(a), the resonance frequency is shifted by $-11$ ($9$) kHz, whereas for the blue-blue sideband, in Fig.~\ref{Lambdips}(c), the shift is $7$ ($13$) kHz. 
The resonance frequency of the Lamb dip is identical, within uncertainty limits, and positioned at the blue marker lines in Fig.~\ref{Lambdips}. 
This shows, within error margins, that in all three configurations the same single Lamb dip is measured.

\begin{figure}[hb]
\begin{center}
\includegraphics[width=\linewidth]{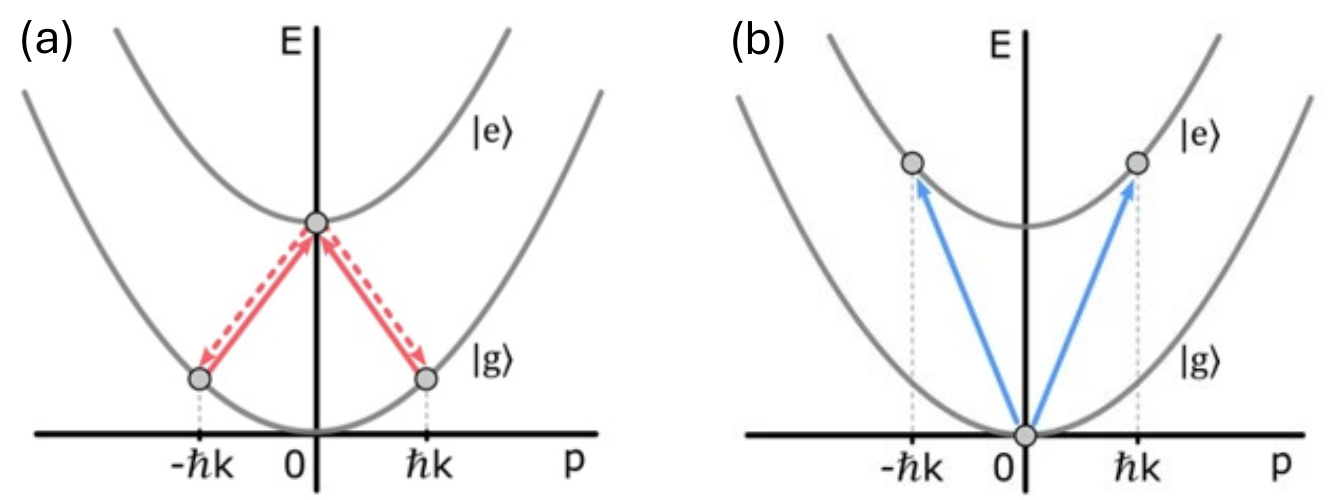}
\caption{\label{Recoil_model}
Curves displaying the sum of internal energy of ground and excited states, |g$\rangle$ and |e$\rangle$, augmented with kinetic energy $\hbar^2 k^2/2M$.
(a) The red recoil component is produced from molecules moving at momentum $p = \pm\hbar k$, pumped into the excited state (solid arrow) and decaying by stimulated emission to the ground state at momentum $p=0$. (b) The blue recoil component is produced from ground state molecules at $p=0$, excited upon absorption into the excited state moving at $p=\pm \hbar k$. 
}
\end{center}
\end{figure}

As discussed previously~\cite{Cozijn2023}, the measured line shape excludes the occurrence of an underlying recoil doublet, as was found in the only reported cases of saturated molecular absorption, where the resolution was sufficient to resolve the recoil structure~\cite{Hall1976b,Bagayev1991}. 
In the case of the H$_2$ S(0) (2-0) line the recoil shift amounts to $h\nu_0^2 / 2Mc^2$ = 70 kHz ($M$ the mass of the H$_2$ molecule), and the splitting of the recoil doublet is 140 kHz.
The blue recoil component can be interpreted as resulting from an absorption process, where excess energy is required to populate the excited state. 
The red recoil component relates to a process of stimulated emission (see Fig.~\ref{Recoil_model}), and was only observed at equal strength if the transition is fully saturated, like in the case of the methane lines at 3.39 $\mu$m~\cite{Hall1976b,Bagayev1991}.

The absence of the red-shifted recoil component can be attributed to the low degree of saturation under such conditions. In a simple two-level model, the fraction of molecules in the excited state is given by 
\begin{equation}
    n_{e} = \frac{s/2}{s+1},
\end{equation}
with the saturation parameter $s$ defined as
\begin{equation}
    s= \frac{2}{\pi} \frac{S_{ge}I c}{\Gamma^{2}hf} .
\end{equation}
Here, $S_{ge} = 3.0\times 10^{-27}$ cm/molecule is the line intensity for the S(0) (2-0) transition in H$_{2}$ at 72~K, computed from the room-temperature experimental result of $1.56 \times 10^{-27}$ cm/molecule~\cite{Kassi2014,Fleurbaey2023}. We assume that no ortho-para conversion takes place~\cite{Stankiewicz2025}, since on every measurement day a fresh H$_2$ sample is taken upon evacuation of the cavity.
$I=3.25\times10^{8}$~W/m$^{2}$ is the laser power density corresponding to 150~W of circulating power at a beam waist of $w_{0}=0.54$~mm, $h$ and $c$ are the Planck constant and speed of light, respectively, and $f$ is the transition frequency. 
The full-width at half-maximum, $\Gamma$ (approximately 250~kHz) is dominated by transit-time broadening, which can be estimated as
\begin{equation}
    \Gamma \approx \frac{1}{2\pi} \frac{v}{2w_{0}},
\end{equation}
with the most probable speed of H$_{2}$ molecules $v=770$~m/s at 72~K, yielding a value of 226~kHz, that agrees reasonably well with the measured FWHM. Substituting these values into the expression for $s$ and $n_{e}$ leads to an excited-state population of $n_{e}\approx 3.5\times10^{-4}$.

This low excited state population obstructs the process presented in Fig~\ref{Recoil_model}(a)
and it explains why in the Lamb dip experiments of the H$_2$ S(0) (2-0) quadrupole line only a single recoil component is observed, which should be interpreted as the blue recoil component represented by Fig~\ref{Recoil_model}(b).
This also means that the zero-recoil position, corresponding to the difference in the ($v=2,J=2$) and ($v=0,J=0$) quantum levels in H$_2$, is at $252\,016\,361\,159\,(6)$ kHz.

\vspace{0.2cm}

\begin{figure}[htb]
\begin{center}
\includegraphics[width=1.0\linewidth]{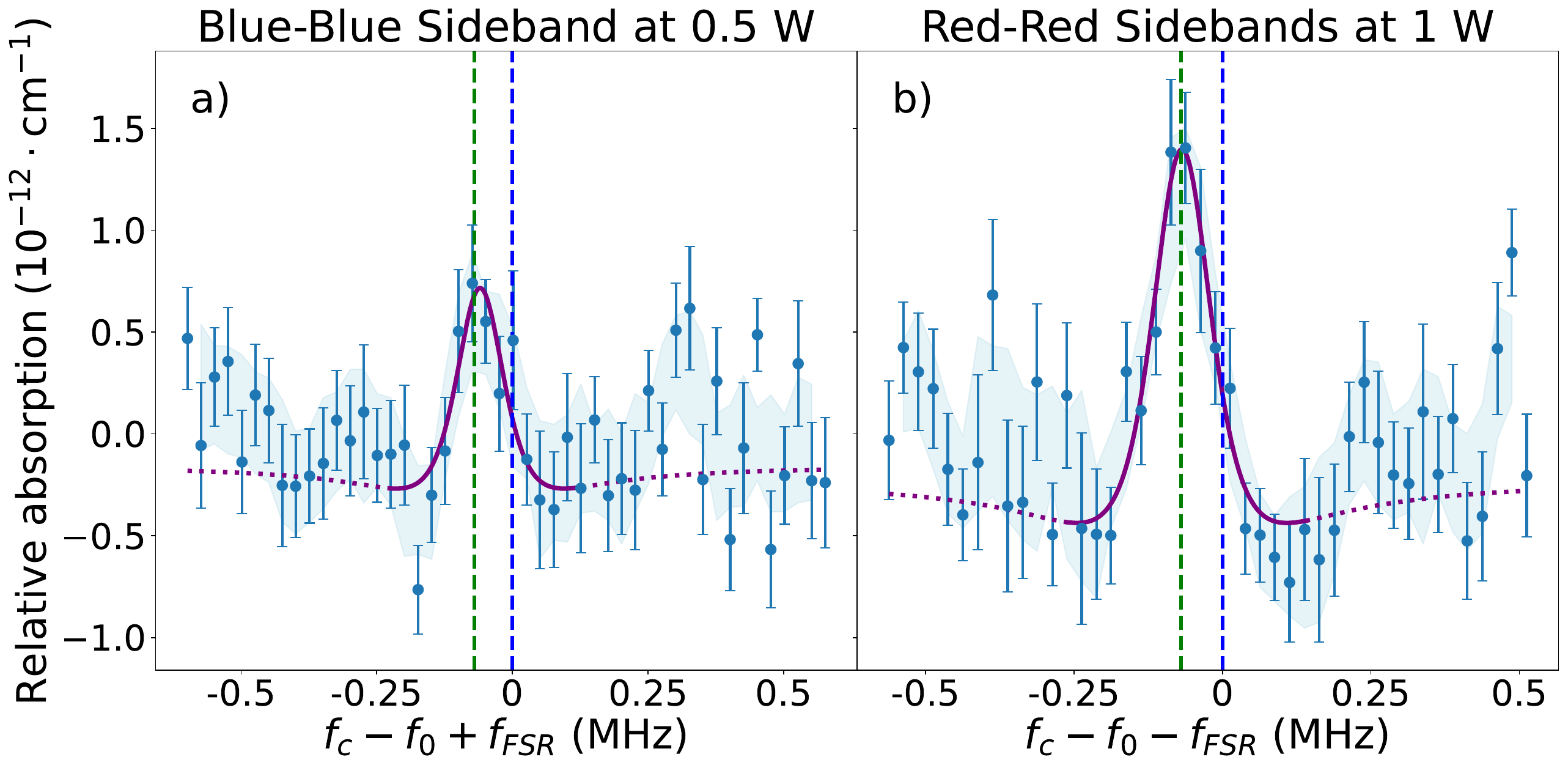}
\caption{\label{Linear}
Linear absorption spectra of the  S(0) (2-0) quadrupole line of H$_2$ recorded at low powers at resonance at $p=$ 0.5 Pa, $T=72$ K, with a trapping field of 2 kW shifted by one FSR (405 MHz). (a) The blue-blue sideband absorption feature at 0.5 W power; (b) The red-red sideband feature at 1 W. In both panels the blue-dashed marker represents the Lamb-dip blue recoil position, and the green-dashed marker represents the zero recoil location. A 3-point moving average interpolation confidence band is added for clarity due to the low signal-to-noise ratio and the absence of a quantitative lineshape model. The ordinary $1f$ NICE-OHMS lineshape model is only fitted in the range of the solid line.
}
\end{center}
\end{figure}

After the presentation of results of the Lamb dips of saturation spectroscopy in the above we alter the experimental conditions. 
The optical configurations as displayed in the red-red and blue-blue sideband schemes of Fig.~\ref{Sideband-scheme} are retained, but intracavity circulating powers are changed. 
The power at the $f_c$ carrier frequency is kept at 2 kW, while the modulation of the electro-optic modulator is reduced to have only $\leq 1$ W power in the sidebands at frequencies $f_c \pm f_{\rm FSR}$. The S(0) (2-0) resonance is now probed in the region of the red- and blue sidebands, with a power at resonance far too low to saturate the transition. 
The observed spectra, obtained after some 14 hours of data averaging, are displayed in Fig.~\ref{Linear}.
A first characteristic is the amplitude sign reversal of the resonances, when compared to the Lamb-dip features in Fig.~\ref{Lambdips}, meaning that the resonances in Fig.~\ref{Linear} are observed in direct absorption.
Note that a Lamb dip signifies a reduced absorption signal.

A second characteristic of the H$_2$ absorption resonances in Fig.~\ref{Linear} is in the extremely narrow width, much narrower than the Doppler profile at the set temperature of 72 K, which is  on the order of 1 GHz, see Fig~\ref{Sideband-scheme}(d). 
These absorption resonances, four orders of magnitude narrower than the Doppler profile, exhibit widths of $\ sim 250$ kHz, similar to the Lamb dips in Fig.~\ref{Lambdips}.
This similarity in width indicates that in both the Lamb-dip spectra and in the linear absorption spectra molecules moving in the plane perpendicular to the laser beam are probed and, hence, the line width in both cases is limited by the transient-time broadening.
Hence, we conclude that in the case of linear absorption the molecules are confined in a one-dimensional (1D) optical lattice produced by the intracavity standing wave of the carrier at $f_c$, as was suggested by Letokhov and Chebotayev~\cite{LetokhovChebotayev}.

As for a quantitative analysis a trapping power at the detuned carrier frequency of 2~kW at a waist of $\omega_0=0.54$~mm for back-and-forth traveling beams corresponds to an intensity of $I=4.33$~kW/mm$^2$.
The dipole polarizability of H$_2$ in its ground rovibrational state at 1189~nm is $\langle\alpha_{\nu}\rangle=5.45$ a.u. or  $\langle\alpha_{\nu}\rangle=8.98 \times 10^{-41}$ C$^2$m$^2$J$^{-1}$, only slightly (0.5\%) higher than the DC value~\cite{Raj2018,Raj2020}.  
The polarizability correction due to the closest quadrupole rovibrational line is only $7 \times 10^{-5}$ a.u. and therewith negligible~\cite{Jozwiak2022}.
The trap depth amounts to
\begin{equation}
    U_0 = \frac{I}{2 \epsilon c} \langle\alpha_{\nu}\rangle = 5.3 \, \mu \mathrm{K}\times k_{\mathrm{B}}.
\end{equation}
The one-dimensional trapping potential along the beam propagation ($y$-axis) represented as $U=U_0 \cos^2 (kx)$ would yield an energy ladder of bound kinetic states~\cite{Monroe1995}
\begin{equation}
    E_n = h\nu(n+\tfrac{1}{2}) = E_{\rm tr} \sqrt{2U_0/mc^2}(n+\tfrac{1}{2}),
    \label{eq:ladder}
\end{equation}
where $E_{\rm tr}$ is the trapping energy of the photon at 1189~nm. 
The energy of the lowest trapped state ($n=0$) then
corresponds to 4.23 $\mu$K. 
Hence, a situation is encountered that just a single kinetic level is bound.
This implies that the trap is far from harmonic and the values obtained only have some approximate or indicative value.

A third characteristic of the observed resonances pertains to their exact frequency. 
The central resonance frequencies of the linear absorption features of Fig.~\ref{Linear} are shifted by -68 (6) kHz for the red-red sideband and -60 (10) kHz for the blue-blue sideband.
Note that for the fitting of these resonances only the central part of the signal is used.
Again, the position of the (blue-shifted recoil) Lamb dip is indicated by the vertical blue-(dashed) lines.
From this we conclude that the absorption resonances observed in Fig.~\ref{Linear} are at the position of zero recoil, as predicted by Letokhov and Chebotayev~\cite{LetokhovChebotayev}.
This phenomenon of linear absorption at the zero-recoil position is considered as a proof for 1D-trapping of H$_2$ molecules by the off-resonant laser field of 2 kW, detuned by 405 MHz. 
This trapping occurs along the propagation direction of the laser beams inside the cavity, while the particles keep their freedom to move perpendicularly.
The trapping occurs in the intensity maxima of the trapping field because the dipole polarizability has a positive sign ($\alpha>0$), as the strong dipole transitions are in the vacuum ultraviolet part of the spectrum.

The narrow absorption resonance is formed from the H$_2$ population having $|v_y|$ not larger than 0.21~m/s (corresponding to the $y$-component kinetic energy of $5.3$~$\mu$K~$k_B$). 
Minima are predicted to occur at both sides of the trapping resonance~\cite{Letokhov1968}, because population is gathered by the trapping effect, and moved to smaller velocities $v_y$.
A velocity of $v_y=0.21$ m/s corresponds to a Doppler frequency shift of 176~kHz towards both sides along the $y$-(laser propagation) axis, hence, the minima are expected to span the $\pm$176~kHz range around the line center. 
These minima are produced by the trapped molecules below $5.3$~$\mu$K amounting to a Boltzmann fraction $P_b=3 \times 10^{-4}$ of the amount of molecules in the sample at 72 K.
The $\pm$176~kHz range is comparable with our transit-time broadening of 226~kHz, hence we do not expect to see as clear minima structure as predicted in Ref.~\cite{Letokhov1968}. 
In Fig.~\ref{Linear}(a) some indication of these minima is apparent.

The dipole force giving rise to trapping should be accompanied by an AC-Stark shift effect. 
Letokhov and Chebotayev argued that molecules in vibrational levels of the same electronic configuration exhibit almost the same Stark shift in a non-resonant light field~\cite{LetokhovChebotayev}. 
For H$_2$ the governing parameters are known.
The mean electric dipole polarizability of the ($v=2,J=2$)  state is 6.45 a.u., slightly higher than that of the (0,0) ground state~\cite{Raj2018,Raj2020}. The effective polarizability for a given state is:
\begin{equation}
\label{eq:AC_Stark}
    \langle\alpha\rangle_{vJM_J} =  \langle\alpha\rangle_{vJ} - \tfrac{2}{3} \langle\gamma\rangle_{vJ} \frac{3M_J^2 - J(J+1)}{(2J-1)(2J+3)}
\end{equation}
with the polarizability anisotropy for the excited state at $\langle\gamma\rangle_{2,2}= 3.07$ a.u.
This leads to a differential polarizability and an AC-Stark shift depending on the $M_J$-value. 
For the linear polarization in the experiment, only the $|\textcolor{red}{\Delta}M_{J}| = 1$ component of the S(0) (2-0) transition is driven. The ground state involves only $M_{J}=0$ hence $\Delta M_{J}$ is equivalent to $M_{J}$ of the excited state. Therefore, we take \eqref{eq:AC_Stark} for $M_{J}=\pm 1$, yielding a predicted AC Stark shift of $-27$ kHz. For details, see the End Matter section.
As discussed above the linear absorption features of Fig.~\ref{Linear} are shifted by less than 10 kHz from the accurately determined zero-recoil position, showing that the 
observed AC-Stark shift is actually very small. 
Similarly, the Lamb dips in both sideband spectra of Fig.~\ref{Lambdips}, where the 2 kW standing wave is also present, underwent an equally low AC-Stark shift below 10 kHz.
The discrepancy between computed and observed AC-Stark shift may be partially explained due to the spatially varying phase between the carrier and sideband standing waves. 
Their intensity antinodes align near the cavity mirrors but are out of phase in the center, meaning the spectroscopic probe is interacting with molecules under different Stark-shifting intensities along the cavity. The resulting signal is a composite, which reduces the net observed shift and may broaden the feature.

The volume of the intracavity radiation field $V=\pi \omega_0^2 L$ amounts to $3.5 \times 10^{-7}$ m$^3$. Assuming that a 1/3 fraction of the molecules is in the high intensity trapping zone and at a pressure of 0.5 Pa at $T=72$ K, with a fraction of 0.25 of the H$_2$ molecules in the lowest $J=0$ para state, this yields a number of $1.5 \times 10^{13}$ particles of which an approximate fraction of $3 \times 10^{-4}$ has a velocity below 0.21 m/s, and will be trapped.
This results in $N_{\rm tr,calc}=4 \times 10^9$.
In view of the cavity length of 37 cm and the number of nodes contained along its axis ($6.2 \times 10^{5}$) the number of trapped molecules per trapping site can be determined at $N_{\rm m} \approx 7000$.

This theoretical estimate can be cross-checked by working backward from the observed absorption signal strength which amounts to roughly $1.5 \times 10^{-12}$~\wn~for the red-red sideband signal (see Fig.~\ref{Linear}).  
Taking into account the partial spatial overlap between the probing sideband and the molecules trapped by the carrier field, a crude estimate yields $N_{\rm tr,obs} = 1 \times 10^9$ for H$_2$ molecules in $J=0$.
Considering that collisions may further reduce the number of trapped molecules this can be considered fair agreement with $N_{\rm tr,calc}$.
The trapping field also captures H$_2$ molecules in ortho-states, which are not detected in absorption.

\vspace{0.1cm}

In conclusion, a proposed one-dimensional trapping scheme, put forward by Letokhov and Chebotayev~\cite{LetokhovChebotayev}, has been experimentally demonstrated for the S(0) (2-0) quadrupole overtone transition in H$_2$.
The observed spectroscopic features reproduce the predictions made by Letokhov and Chebotayev in their Fig.~6.11a~\cite{LetokhovChebotayev} in all aspects: (i) the sign reversal from Lamb dip to linear absorption, (ii) an extremely narrow line width showing that only molecules moving perpendicularly to the laser beam (and being 1D-trapped) contribute to the signal, (iii) the line center is moved to the zero recoil position. 
While such trapping phenomena are expected to occur in all molecules  the 1189~nm transition in the lightest molecule was chosen because it allows for unequivocally distinguishing a line shift between the zero-recoil position and the recoil-affected position. Due to the small mass of H$_2$ the recoil shift becomes exceptionally large with a value of 70 kHz.

Finally, the observation of the resonance at the zero-recoil position proves that in the Lamb-dip experiments of a very weakly saturated transition only a single blue-shifted recoil components is observed. This in itself may impact other studies of weakly saturated molecular spectroscopy, where recoils cannot be distinguished as clearly as in the low-mass H$_2$ molecule. 
This may impact transition frequencies obtained in precision studies of HD~\cite{Tao2018,Cozijn2018}, HT~\cite{Cozijn2024} and H$_2$O~\cite{Tobias2020}.
With the advances in cavity-enhanced spectroscopy combined with frequency metrology techniques~\cite{Gianfrani2024} this 1D-trapping effect may become part of the toolbox of precision molecular spectroscopy and sensing.

\vspace{0.5cm}

This work was part of the 23FUN04 COMOMET project that has received funding from the European Partnership on Metrology, co-financed by the European Union’s Horizon Europe Research and Innovation Programme and from the Participating States (Funder ID: 10.13039/100019599). This research was also funded by the European Union (Grant No 101075678, ERC-2022-STG, H2TRAP).

%

\section*{End matter}
The S(0) 2-0 transition is a quadrupole transition; hence, the laser-molecule interaction does not probe only the local electric field vector but also the gradients of its components. Therefore, the selection rule does not involve only the relative orientation of the polarizations of the two lasers but also the direction of the probe laser propagation. The quantization axis ($z$-axis) is set by the polarization vector of the strong off-resonant laser (which is linear in our case). In our experiment, the probe laser has the same polarization. The spherical components of the electric field gradient $\mathrm{T}^{(2)}_{p}(\nabla \mathbf{E})$ govern the selection rules (see Eqs. (49) in Supp. Mat. to Ref.~\cite{Jozwiak2022}). Out of three spherical components ($p=0$, $\pm 1$, $\pm 2$) only the $p=\pm 1$ term yields a nonzero result. In principle, also the $p=0$ term is nonzero, but the field gradient perpendicular to the laser propagation direction ($\partial_{z}E_{z}$) is much smaller comparing to the gradient along the laser propagation direction ($\partial_{y}E_{z}$); we reiterate that both laser beams propagate along the $y$-axis. This constrains that only $|\Delta M_{J}| = 1$ transitions are allowed. 

\end{document}